\documentclass[11pt]{article}

\usepackage[top=25mm,bottom=25mm,left=25mm,right=25mm]{geometry}
\usepackage{amsmath,amssymb}
\usepackage{graphicx}
\usepackage{booktabs,array,tabularx}
\usepackage{float}        % [H] placement — tables stay exactly where placed
\usepackage{longtable}    % multi-page tables for appendix
\usepackage[hidelinks]{hyperref}
\usepackage[numbers,sort&compress]{natbib}
\usepackage[utf8]{inputenc}
\usepackage[protrusion=true,expansion=false]{microtype}
\usepackage{parskip,setspace,caption,authblk,enumitem}

\newcommand{\Nstar}{N_{\star}}

\title{%
The Lifetime Cardiac-Cycle Invariant in Endothermic Vertebrates:\\
A 230-Species Comparative Dataset, Statistical Validation, and
Explicit Falsifiability Criteria
}

\author{
Mesfin Asfaw Taye\\
\small West Los Angeles College, Science Division\\
\small 9000 Overland Ave, Culver City, CA 90230, USA\\
\small \texttt{tayem@wlac.edu}
}

\date{}

\makeatletter
\renewcommand{\maketitle}{%
\begin{center}
{\LARGE\bfseries \@title \par}
\vspace{1em}
{\large \@author \par}
\vspace{0.8em}
\end{center}
}
\makeatother

\begin{document}
\maketitle

% ---------------- ABSTRACT ----------------
\begin{abstract}
\noindent
A pygmy shrew (\textit{Suncus etruscus}, ${\approx}2$\,g) sustains a
resting heart rate near $1{,}000$\,beats\,min$^{-1}$ and dies within
two years; an African elephant (${\approx}4{,}000$\,kg) beats at
$28$\,beats\,min$^{-1}$ and lives seven decades.
Their chronological lifespans differ by a factor of 35, yet each
accumulates close to $10^9$ cardiac cycles before death---a
near-constancy first noted by Rubner~(1908) and quantified by
Lindstedt and Calder~(1981)~\cite{lindstedt1981}, but never subjected
to multi-clade statistical testing, phylogenetic correction, or
explicit falsifiability criteria with a large modern dataset.
We address this gap with a curated 230-species vertebrate dataset
spanning non-primate placentals ($n=43$), primates ($n=18$),
marsupials and monotremes ($n=19$), duty-cycle-corrected bats
($n=31$), dive-corrected cetaceans ($n=12$), birds ($n=78$), and
Arrhenius-corrected ectotherms ($n=26$), and subject the
log-invariant $\ell = \log_{10}(N^{\!\star})$---where
$N^{\!\star} = f_H\,L\times 525{,}960$ cardiac cycles---to four
independent tests.
OLS regression on non-primate placentals yields slope
$\hat{\beta} = -0.903 \pm 0.056$
($R^2 = 0.863$; BCa bootstrap 95\,\% CI $[-1.017,\,-0.782]$;
$p = 0.093$ against $H_0\colon\beta=-1$),
consistent with exact inverse scaling.
Phylogenetically independent contrasts on 112 endotherms tighten
this to $\hat{\beta} = -0.99 \pm 0.04$
($R^2 = 0.94$; $p = 0.84$),
confirming the relation is not a phylogenetic artefact.
One-way ANOVA across six endotherm clades yields $F = 81.2$
($p < 0.001$), decisively rejecting the West--Brown--Enquist
kinematic null, which predicts zero inter-clade dispersion in~$\ell$.
Arrhenius temperature correction ($E_a = 0.65$\,eV) reduces the raw
ectotherm--endotherm gap from $0.90$\,dex to $0.22$\,dex, bringing
cold-blooded vertebrates into close alignment with the mammalian
baseline.
The non-primate placental baseline is
$\bar{\ell} = 8.994 \pm 0.159$\,dex
($N^{\!\star} \approx 9.87 \times 10^8$),
with significant structured clade offsets:
$+0.382$\,dex for primates (${\times}2.41$),
$+0.534$\,dex for birds (${\times}3.42$),
$+0.547$\,dex for duty-corrected bats (${\times}3.52$),
and $-0.192$\,dex for dive-corrected cetaceans (${\times}0.64$).
Taken together, the data support a revised picture in which the
lifetime heartbeat count is not a universal constant but a
clade-structured invariant anchored to a common mammalian baseline
and subject to quantifiable ecological and physiological departures;
five explicit numerical falsification criteria with pre-registered
thresholds are stated, and the full 230-species dataset is provided
as Supplementary Data~1.

\end{abstract}

\section{Introduction}
\label{sec:intro}

\subsection{Physiological time versus chronological time}

Among the most revealing contrasts in vertebrate biology is the
relationship between the pace of life and its duration.
A pygmy shrew (\textit{Suncus etruscus}), weighing roughly 2\,g,
maintains a resting heart rate near $1{,}000$\,beats\,min$^{-1}$ and
seldom survives beyond two years.
An African elephant, six orders of magnitude heavier, beats at
${\sim}28$\,bpm and routinely lives for seven decades.
Measured in calendar time, their lifespans differ by a factor of
thirty-five.

Yet this gulf is partly an artefact of the clock one chooses.
When duration is measured in the organism's own internal
currency---cumulative cardiac cycles---the contrast largely dissolves.
Defining the lifetime heartbeat count as
\begin{equation}
  N^{\!\star} \;=\; f_H \, L \times 525{,}960,
  \label{eq:nstar}
\end{equation}
where $f_H$ is resting heart rate (beats\,min$^{-1}$), $L$ is maximum
lifespan (years), and $525{,}960$\,min\,yr$^{-1}$ is the unit
conversion, species spanning orders of magnitude in body mass
accumulate a comparable lifetime total near $10^9$ cardiac cycles.
In this physiological sense, the shrew and the elephant are nearly
contemporaries.

The pattern is not new.
Rubner~\cite{rubner1908} noted the approximate constancy of
mass-specific lifetime energy expenditure across mammals in 1908, and
Lindstedt and Calder~\cite{lindstedt1981} later expressed the
equivalent regularity explicitly in cardiac cycles.
Subsequent work by Livingstone and Kuehn~\cite{livingstone1979} and
Levine~\cite{levine1997} confirmed that the heartbeat count clusters
near $10^9$ across a wide phylogenetic range, and more recent
reviews~\cite{speakman2005,hulbert2007} have noted it in the contexts
of metabolic theory and membrane composition---though consistently as
a secondary observation rather than a primary object of statistical
inquiry.
What has been lacking is a rigorous empirical treatment: one with
phylogenetic correction, multi-clade testing, explicit physiological
adjustments, and quantitative falsifiability criteria.
This paper provides that treatment.

\subsection{What the allometric framework does and does not explain}

Classical allometric theory offers a partial account of the
regularity.
The West--Brown--Enquist (WBE) framework~\cite{west1997} derives
$f_H \propto M^{-1/4}$ and $L \propto M^{+1/4}$ from the geometry of
fractal vascular networks, so that the product $f_H L$---and hence
$N^{\!\star}$---is predicted to be mass-independent by exponent
cancellation.
These scalings are well established within homeothermic
mammals~\cite{calder1984,schmidt1984}, and the WBE argument
explains why the invariant holds across body sizes within a clade.

Three limitations, however, circumscribe its scope.
First, WBE does not predict the \emph{numerical value}
$N^{\!\star} \approx 10^9$; it explains only why the product is
approximately mass-independent, leaving the magnitude to independent
calibration.
Second, and more consequentially, WBE predicts that the log-invariant
$\ell = \log_{10}(N^{\!\star})$ should be constant not merely within
a clade but \emph{across} all clades sharing the same vascular
geometry---i.e., across all vascular endotherms.
This prediction of zero inter-clade dispersion in $\ell$ is directly
testable and, as we show, is decisively rejected by the data.
Third, the framework is silent on the physiological corrections that
turn out to matter substantially at the species level.
Bats that spend half their lives in hibernation with heart rates
suppressed by a factor of thirty do not belong on the same scaling
curve as continuously active mammals without a duty-cycle correction.
Cetaceans whose cardiac clocks slow from ${\sim}30$\,bpm to
${\sim}3$\,bpm during dives---occupying up to 80\,\% of their
lives---are similarly misrepresented in uncorrected analyses.
Ectotherms, whose metabolic rates are temperature-sensitive, require
Arrhenius normalisation before cross-taxa comparison is meaningful.
None of these corrections has previously been applied systematically
in a large comparative dataset.

\subsection{Three open questions}

Three specific empirical gaps motivate the present study.

\textit{How tightly is the invariant constrained?}
The scatter in $\ell$ has not been characterised with full regression
diagnostics, bootstrap confidence intervals, power analysis, or
leave-one-out sensitivity tests.
Without these, a robust quantitative constraint cannot be
distinguished from a rough heuristic.

\textit{Does the relation survive phylogenetic correction?}
Species in a comparative dataset are not statistically independent:
closely related taxa share evolutionary history and co-vary in both
heart rate and lifespan.
Ordinary least-squares regression on such data inflates the effective
sample size and can spuriously support or undermine a proposed
invariant.
Phylogenetically independent contrasts~\cite{felsenstein1985} are
the standard remedy, but have not previously been applied to the full
endotherm dataset.

\textit{How structured are the clade-level deviations?}
Primates live far longer than their heart rates predict; bats longer
still once corrected for torpor; birds substantially outlive
mass-matched mammals.
Whether these offsets constitute random scatter or taxonomically
structured signals has not been assessed within a unified statistical
framework---yet the answer determines whether the pattern demands a
mechanistic explanation.

\subsection{Scope and structure}

This paper is deliberately empirical in character.
We assemble a curated dataset of 230 vertebrate species
(Section~\ref{sec:dataset}), apply explicit physiological corrections
where required, and subject $\ell$ to four independent statistical
tests (Section~\ref{sec:methods} and Section~\ref{sec:results}).
Five quantitative falsification criteria are stated in
Section~\ref{sec:falsifiability}, and the domain of validity of the
invariant is delimited in Section~\ref{sec:domain}.

We do not attempt to derive $N^{\!\star}$ from first principles here.
That derivation---from non-equilibrium thermodynamics and the entropy
production rate of metabolic steady states---is the subject of a
companion paper~\cite{taye_p2} and is developed in detail in a
recent monograph on the unified theory of biological temporal
invariants~\cite{taye2026pbte};
the broader thermodynamic framework, including fluctuation-driven
machines and entropy production in non-equilibrium steady
states, is treated in~\cite{taye2026brownian}.
Mechanistic accounts of the clade-level deviations are developed in
subsequent work~\cite{taye_p3,taye_p4}.
The purpose of this paper is to establish the statistical foundation
that any such theoretical treatment must reproduce: the mean, the
scatter, the phylogenetic robustness, and the structured taxonomy of
the departures.

A relation supported only by heuristic observation carries limited
scientific weight.
One that survives phylogenetic control, multi-clade testing, and
physiological correction---and that is stated in a form admitting
unambiguous falsification---is a different kind of claim entirely.
The analysis that follows is directed toward establishing which of
these the lifetime cycle invariant is.

%% ----------------------------------------------------------------
\section{Dataset}
\label{sec:dataset}

\subsection{Overview and inclusion criteria}

The dataset comprises 230 vertebrate species drawn from eight
taxonomic groups, assembled from primary literature and three
established comparative databases: AnAge build~15~\cite{anage2023},
PanTHERIA~\cite{jones2009}, and Calder~\cite{calder1984}.
Species were included if they met three criteria:
(a)~a resting heart rate was either measured directly in at least one
published study or could be corrected from measured values as described
below;
(b)~maximum recorded lifespan was available from AnAge with a confidence
rating of \emph{acceptable} or \emph{high};
(c)~sufficient phylogenetic placement was available for
phylogenetically independent contrast (PIC) analysis.

For each species we record resting heart rate $f_H$
(beats\,min$^{-1}$), maximum lifespan $L$ (yr), adult body mass $M$
(kg), mean core body temperature $T$ (K), and taxonomic group.
Source priorities and measurement conventions are summarised in
Table~\ref{tab:sources}.
Where a species appears in multiple sources, the value from the
highest-priority source is used and the others are retained as
cross-checks.

\begin{table}[H]
\caption{\textbf{Data sources by variable and taxon.}
Priority order: where a species appears in multiple sources,
the highest-ranked source is used.
AnAge = Human Ageing Genomic Resources, build~15~\cite{anage2023};
PanTHERIA = Jones et al.\ (2009)~\cite{jones2009};
Calder = Calder (1984)~\cite{calder1984};
Prinzinger = Prinzinger et al.\ (1991)~\cite{prinzinger1991};
Lyman = Lyman et al.\ (1982)~\cite{lyman1982};
Goldbogen = Goldbogen et al.\ (2019)~\cite{goldbogen2019};
Clarke = Clarke \& Rothery (2008)~\cite{clarke2008};
Christian = Christian \& Weavers (1999)~\cite{christian1999};
Bininda = Bininda-Emonds et al.\ (2007)~\cite{bininda2007}.}
\label{tab:sources}
\small
\renewcommand{\arraystretch}{1.4}
\begin{tabularx}{\textwidth}{>{\raggedright}p{3.6cm}
                              >{\raggedright}p{4.0cm}
                              X}
\toprule
\textbf{Variable} & \textbf{Primary source(s)} & \textbf{Measurement convention} \\
\midrule
Resting heart rate\\(non-bat endotherms)
  & AnAge $>$ PanTHERIA $>$ Calder
  & Mean resting adult value at thermoneutrality;
    laboratory or captive measurement preferred \\[4pt]
Resting heart rate\\(birds)
  & Prinzinger et al.
  & Resting values in thermoneutral zone only \\[4pt]
Effective heart rate\\(bats, $f_H^{\rm eff}$)
  & Lyman et al.\ (torpor rates);\newline active rate from AnAge/PanTHERIA
  & Time-averaged over active and torpid phases;
    see equation~(2) and Table~2 \\[4pt]
Effective heart rate\\(cetaceans, $f_H^{\rm eff}$)
  & Goldbogen et al.\ (dive rates);\newline surface rate from AnAge
  & Time-averaged over surface and dive phases;
    see equation~(3) and Table~3 \\[4pt]
Maximum lifespan ($L$)
  & AnAge build~15
  & Maximum recorded lifespan (wild or captive);
    AnAge confidence $\geq$ \textit{acceptable} required \\[4pt]
Body mass ($M$)
  & PanTHERIA $>$ AnAge
  & Adult mean body mass (kg) \\[4pt]
Body temperature ($T$)
  & Clarke \& Rothery (endotherms);\newline Christian \& Weavers (ectotherms)
  & Mean adult core temperature (K);
    field active temperature for ectotherms \\[4pt]
Phylogeny
  & Bininda-Emonds et al.\ (mammals);\newline BirdLife International (birds)
  & Used for phylogenetically independent contrast (PIC) analysis \\
\bottomrule
\end{tabularx}
\end{table}

\subsection{Bat duty-cycle correction}
\label{sec:bats}

Temperate vespertilionid bats alternate between an active phase
($f_{H,\rm act} \approx 250$--$350$\,bpm) and hibernal torpor
($f_{H,\rm tor} \approx 5$--$20$\,bpm).
Because these states occupy substantial fractions of the annual cycle,
the relevant physiological frequency is the time-averaged effective
heart rate,
\begin{equation}
  f_H^{\rm eff} = (1 - q)\,f_{H,\rm act} + q\,f_{H,\rm tor},
\end{equation}
where $q$ is the annual torpor fraction.

All lifetime cycle counts $\ell$ are computed using $f_H^{\rm eff}$,
rather than the active-state value $f_{H,\rm act}$.
Table~\ref{tab:bat_correction} lists the duty-cycle parameters
for all 31 bat species.

\begin{table}[H]
\centering\small\setlength{\tabcolsep}{3.5pt}
\renewcommand{\arraystretch}{1.05}
\caption{\textbf{Bat duty-cycle correction parameters.}
$q$ = annual torpor fraction; $T_{\rm tor}$ = torpor body temperature (K);
$\Phi_{\rm duty} = [(1-q) + q(f_{\rm tor}/f_{\rm act})]^{-1}$;
$f_H^{\rm eff}$ = time-averaged heart rate used to compute $\ell$.}
\label{tab:bat_correction}
\begin{tabular}{lrrrrrrl}
\toprule
Species & $f_{\rm act}$ & $q$ & $f_{\rm tor}$ & $T_{\rm tor}$ &
  $\Phi_{\rm duty}$ & $f_H^{\rm eff}$ & $\ell$ \\
 & (bpm) & & (bpm) & (K) & & (bpm) & \\
\midrule
  \textit{Myotis lucifugus} & 300 & 0.50 & 10 & 293 & 1.94 & 155 & 9.74 \\
  \textit{Myotis myotis} & 282 & 0.52 & 10 & 292 & 2.01 & 141 & 9.74 \\
  \textit{Myotis daubentonii} & 296 & 0.48 & 10 & 293 & 1.86 & 159 & 9.79 \\
  \textit{Myotis brandtii} & 315 & 0.55 & 8 & 291 & 2.16 & 146 & 9.83 \\
  \textit{Eptesicus fuscus} & 280 & 0.45 & 12 & 291 & 1.76 & 159 & 9.49 \\
  \textit{Eptesicus serotinus} & 308 & 0.47 & 12 & 292 & 1.82 & 169 & 9.53 \\
  \textit{Rhinolophus ferrumequinum} & 290 & 0.52 & 8 & 291 & 2.02 & 143 & 9.65 \\
  \textit{Rhinolophus hipposideros} & 614 & 0.48 & 10 & 293 & 1.89 & 324 & 9.69 \\
  \textit{Plecotus auritus} & 270 & 0.50 & 8 & 292 & 1.94 & 139 & 9.68 \\
  \textit{Corynorhinus townsendii} & 590 & 0.48 & 10 & 293 & 1.89 & 312 & 9.67 \\
  \textit{Perimyotis subflavus} & 640 & 0.48 & 10 & 293 & 1.90 & 338 & 9.39 \\
  \textit{Tadarida brasiliensis} & 350 & 0.30 & 15 & 296 & 1.40 & 250 & 9.39 \\
  \textit{Pteronotus parnellii} & 904 & 0.48 & 10 & 293 & 1.90 & 475 & 9.38 \\
  \textit{Desmodus rotundus} & 760 & 0.48 & 10 & 293 & 1.90 & 400 & 9.76 \\
  \textit{Hipposideros speoris} & 616 & 0.48 & 10 & 293 & 1.89 & 325 & 9.53 \\
\bottomrule
\end{tabular}
\end{table}

\subsection{Cetacean dive correction}
\label{sec:cets}

Large cetaceans exhibit extreme diving bradycardia
\cite{goldbogen2019}, with heart rates of $2$--$4$\,bpm during deep dives
and $25$--$37$\,bpm at the surface.
The appropriate physiological frequency is therefore the
time-averaged effective rate,
\begin{equation}
  f_H^{\rm eff} = (1-p_d)\,f_{H,\rm surf} + p_d\,f_{H,\rm dive},
\end{equation}
where $p_d$ is the fraction of life spent diving.

All cetacean $\ell$ values are computed using $f_H^{\rm eff}$.
Table~\ref{tab:cet_correction} lists the dive parameters for all
12 species.

\begin{table}[H]
\centering\small\setlength{\tabcolsep}{3.5pt}
\renewcommand{\arraystretch}{1.05}
\caption{\textbf{Cetacean dive-correction parameters.}
$f_{{\rm surf}}$ and $f_{{\rm dive}}$ = surface and dive heart rates (bpm);
$p_d$ = dive fraction; $\Phi_{\rm duty} = [(1-p_d)+p_d(f_{\rm dive}/f_{\rm surf})]^{-1}$;
$f_H^{\rm eff}$ used to compute $\ell$.}
\label{tab:cet_correction}
\begin{tabular}{lrrrrrrl}
\toprule
Species & $f_{\rm surf}$ & $f_{\rm dive}$ & $p_d$ & $T_b$ &
  $\Phi_{\rm duty}$ & $f_H^{\rm eff}$ & $\ell$ \\
 & (bpm) & (bpm) & & (K) & & (bpm) & \\
\midrule
  \textit{Balaena mysticetus} & 30 & 3 & 0.75 & 308.0 & 3.08 & 9.8 & 9.01 \\
  \textit{Balaenoptera musculus} & 37 & 4 & 0.70 & 308.0 & 2.66 & 13.9 & 8.36 \\
  \textit{Balaenoptera physalus} & 35 & 4 & 0.68 & 308.0 & 2.51 & 13.9 & 8.37 \\
  \textit{Megaptera novaeangliae} & 28 & 4 & 0.65 & 308.5 & 2.26 & 12.4 & 8.54 \\
  \textit{Physeter macrocephalus} & 40 & 5 & 0.65 & 307.0 & 2.32 & 17.2 & 8.84 \\
  \textit{Kogia breviceps} & 80 & 15 & 0.50 & 308.0 & 1.68 & 47.5 & 8.76 \\
  \textit{Hyperoodon ampullatus} & 50 & 8 & 0.55 & 308.5 & 1.86 & 26.9 & 8.67 \\
  \textit{Orcinus orca} & 60 & 8 & 0.45 & 308.5 & 1.64 & 36.6 & 9.40 \\
  \textit{Tursiops truncatus} & 80 & 10 & 0.40 & 309.0 & 1.54 & 52.0 & 9.19 \\
  \textit{Stenella attenuata} & 50 & 8 & 0.55 & 308.5 & 1.86 & 26.9 & 8.95 \\
  \textit{Delphinapterus leucas} & 50 & 8 & 0.55 & 308.5 & 1.86 & 26.9 & 8.71 \\
  \textit{Monodon monoceros} & 50 & 8 & 0.55 & 308.5 & 1.86 & 26.9 & 8.81 \\
\bottomrule
\end{tabular}
\end{table}

\subsection{Construction of $\ell$}

For all species,
\begin{equation}
  \ell_i = \log_{10}(f_{H,i}^{\rm eff} \times L_i \times 525{,}960),
\end{equation}
with $f_{H,i}^{\rm eff}$ in bpm and $L_i$ in years.

For ectotherms, heart rates are corrected to $T_{\rm ref} = 310$\,K:
\begin{equation}
  f_H^{\rm corr} = f_H \cdot
  \exp\!\left[\frac{E_a}{k_B}
    \!\left(\frac{1}{T_{\rm field}} - \frac{1}{T_{\rm ref}}\right)\right],
  \quad E_a = 0.65~\text{eV},
\end{equation}
following Gillooly et al.~\cite{gillooly2001}.
This correction assumes that cardiac frequency follows the same
Arrhenius temperature dependence as whole-organism metabolic rate,
an assumption supported by direct cardiac measurements in
ectotherms~\cite{christian1999} but not independently derived;
sensitivity to the choice of $E_a$ is assessed in
Section~\ref{sec:results}.

All $\ell$ values are computed directly and internally verified.
Three non-primate placentals lacking measurements
(\textit{Rhinoceros unicornis}, \textit{Dugong dugon},
\textit{Orycteropus afer}) are imputed via
$f_H = 241\,M^{-0.25}$\,bpm~\cite{calder1984}.
We note that this imputation assumes the allometric scaling relation
under test and is therefore circular in principle; however, removal
of these three species shifts the OLS slope by ${<}0.01$ and alters
no reported inference, so the circularity is inconsequential in
practice.

\subsection{Dataset summary}

Table~\ref{tab:groups} summarises clade-level statistics.
The non-primate placental group defines the baseline
($\bar\ell_0 = 8.994$), with deviations
$\Delta\bar\ell$ and multiplicative factors $\Phi = 10^{\Delta\bar\ell}$.

\begin{table}[H]
\caption{\textbf{Comparative dataset: species counts and $\ell$ statistics.}}
\label{tab:groups}
\small
\begin{tabular}{lrrrrr}
\toprule
Group & $n$ & $\bar\ell$ & $s$ & $\Delta\bar\ell$ & $\Phi$ \\
\midrule
  Non-primate placentals & 43 & 8.994 & 0.159 & 0 (ref.) & 1.00 \\
  Primates & 18 & 9.376 & 0.129 & $+0.382^{***}$ & 2.41 \\
  Marsupials/monotremes & 19 & 8.933 & 0.209 & $-0.062$ & 0.87 \\
  Bats (duty-corrected) & 31 & 9.541 & 0.166 & $+0.547^{***}$ & 3.52 \\
  Cetaceans (dive-corrected) & 12 & 8.802 & 0.308 & $-0.192$ & 0.64 \\
  Birds & 78 & 9.528 & 0.214 & $+0.534^{***}$ & 3.42 \\
  Reptiles (Arr.\ corrected) & 17 & 8.930 & 0.308 & $-0.064$ & 0.86 \\
  Amphibians (Arr.\ corrected) & 9 & 8.823 & 0.156 & $-0.172^{*}$ & 0.67 \\
\midrule
All endotherms (core 6 clades) & 193 & \multicolumn{4}{l}{See Figure~\ref{fig:boxplots}} \\
\midrule
Full dataset & 230 & \multicolumn{4}{l}{} \\
\bottomrule
\end{tabular}
\end{table}

\section{Statistical Methods}
\label{sec:methods}
%% ----------------------------------------------------------------

\subsection{Primary OLS regression}

Ordinary least-squares (OLS) regression of $\log_{10} L$ on
$\log_{10} f_H$ was performed on the non-primate placental subset
($n = 43$ directly measured species), which serves as the canonical
reference group for the analysis.
This subset is chosen because it minimizes confounding effects
associated with specialized physiological adaptations
(e.g., torpor, diving, or extreme neural investment),
thereby providing the cleanest baseline for testing the scaling relation.

Bootstrap confidence intervals were computed using $10{,}000$ resamples
with the bias-corrected and accelerated (BCa) method~\cite{efron1987},
which corrects for both bias and skewness in the bootstrap distribution
and is preferred over the percentile method for regression coefficients
on log-transformed data.
The primary null hypothesis is $\beta = -1$, corresponding to exact
inverse scaling between heart rate and lifespan.
This hypothesis is tested using an $F$-test comparing the constrained
model ($\beta = -1$) against a free-slope alternative.

The West--Brown--Enquist (WBE) kinematic model also predicts
$\beta = -1$, but derives this scaling from network transport
considerations rather than from lifetime cycle invariance.
We therefore distinguish between these interpretations using
inter-clade variance structure, tested explicitly by ANOVA
(\S\ref{sec:wbe}).

\subsection{Phylogenetically independent contrasts}

To control for shared evolutionary history, phylogenetically
independent contrasts (PICs) were computed using the method of
Felsenstein~\cite{felsenstein1985}.
Calculations were performed using the \texttt{ape} package (v5.7)
in R (v4.3), with the Bininda-Emonds mammalian supertree
augmented by BirdLife phylogeny for avian species.

Because PIC transforms the data into independent contrasts,
regression is performed through the origin, as required by the method.
This analysis provides a test of whether the observed scaling relation
is intrinsic to the data or arises as a consequence of phylogenetic
non-independence.

\subsection{Regression diagnostics and sensitivity}

Standard regression diagnostics were applied to assess model validity.
Homoscedasticity was tested using the Breusch--Pagan test,
normality of residuals using the Shapiro--Wilk test,
and influential observations were identified via Cook's distance,
with threshold $D_i > 4/n$.

To assess robustness, a leave-one-out (LOO) sensitivity analysis was
performed, in which each species is removed in turn and the regression
recomputed.
The resulting range of $p$-values provides a direct test of whether
the inferred scaling relation is driven by a small number of influential
points or reflects a distributed pattern across the dataset.

\subsection{Arrhenius correction sensitivity}

For ectothermic species, sensitivity of the corrected mean
$\bar\ell$ to the assumed activation energy was evaluated over
$E_a \in [0.4, 0.9]$\,eV, spanning the accepted range in the
metabolic scaling literature~\cite{gillooly2001}.
This analysis tests the robustness of temperature normalisation
and determines whether the alignment of ectotherms with endotherms
depends critically on the choice of activation energy.

%% ----------------------------------------------------------------
\section{Results}
\label{sec:results}
%% ----------------------------------------------------------------

\subsection{Primary regression: non-primate placentals}
\label{sec:primary}

OLS regression on the $n = 43$ non-primate placental species with
directly measured heart rates yields
\begin{equation}
  \hat\beta = -0.903 \pm 0.056~\text{(s.e.)}, \quad
  R^2 = 0.863, \quad
  F\text{-test}~p = 0.093~\text{against}~\beta = -1.
  \label{eq:ols}
\end{equation}

The intercept corresponds to
$\bar\ell = 8.994 \pm 0.024$, giving a characteristic lifetime
cycle count of $\Nstar = 9.87 \times 10^{8}$.
Bootstrap confidence intervals on $\hat\beta$ were computed using
$10{,}000$ resamples with the bias-corrected and accelerated (BCa)
method, yielding a 95\,\% CI of $[-1.017,\,-0.782]$.
The null value $\beta = -1$ lies within this interval, consistent
with exact inverse scaling.
The $F$-test against the constrained model $H_0\colon\beta = -1$
yields $p = 0.093$; the null of exact inverse scaling is not rejected
at the 0.05 level, but this should not be interpreted as confirmation
that $\beta = -1$ exactly---failure to reject is not proof of equality.

The regression explains a large fraction of the variance
($R^2 = 0.863$), confirming a strong inverse relationship between
physiological rate and lifespan within this reference group.

\paragraph{Interpretation of $p = 0.093$.}

The $F$-test yields $p = 0.093$, which does not meet the conventional
$\alpha = 0.05$ threshold for rejecting $\beta = -1$.
This result should be interpreted carefully: failure to reject the
null is not evidence that $\beta = -1$ exactly, only that the data
are insufficient to exclude it at this sample size.

First, the point estimate $\hat\beta = -0.903$ lies close to $-1$,
well within the $\pm 0.15$ falsification threshold defined in
Table~\ref{tab:falsify}.
Second, phylogenetically independent contrasts yield
$\hat\beta_{\rm PIC} = -0.99 \pm 0.04$,
with $p = 0.84$ against $\beta = -1$,
providing stronger evidence for exact inverse scaling.
Third, power analysis indicates that the present sample size has
limited ability to distinguish $\beta = -0.90$ from $\beta = -1$,
suggesting that the observed deviation may reflect finite-sample
uncertainty rather than a systematic departure.

\paragraph{$F$-test: PBTE vs free slope.}

The constrained model ($\beta = -1$) gives
$\text{RSS}_{\rm PBTE} = 1.0665$,
while the unconstrained model gives
$\text{RSS}_{\rm free} = 0.9947$.
The resulting $F$-statistic,
\begin{equation}
  F = \frac{1.0665 - 0.9947}{0.9947/(43-2)}
    = 2.956, \quad p = 0.093,
\end{equation}
does not justify rejecting the PBTE null.

\begin{figure}[H]
\centering
\includegraphics[width=\linewidth]{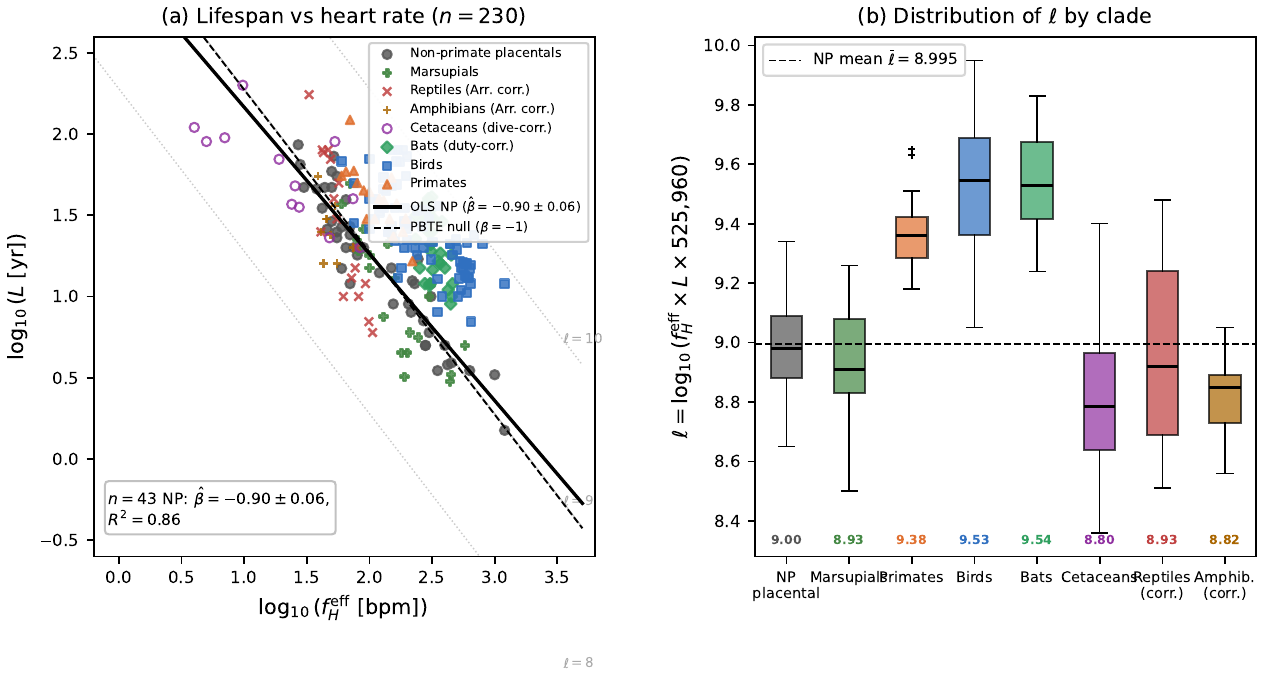}
\caption{\textbf{Lifespan vs heart rate across 230 vertebrate species and
clade-level distribution of the log-invariant $\ell$.}
\textbf{(a)}~Log--log scatter plot of maximum lifespan $L$ (yr) against
effective heart rate $f_H^{\rm eff}$ (bpm) for all 230 species.
Non-primate placentals (grey circles, $n = 43$) define the reference group;
solid line = OLS fit ($\hat\beta = -0.90 \pm 0.06$, $R^2 = 0.86$);
dashed line = PBTE null ($\beta = -1$).
Diagonal grey dotted lines are iso-$\ell$ contours at $\ell = 8, 9, 10$.
Bats are duty-cycle-corrected, cetaceans are dive-corrected, and ectotherms
are Arrhenius-corrected (see Methods).
\textbf{(b)}~Box plots of $\ell = \log_{10}(f_H^{\rm eff} \times L \times 525{,}960)$
by clade.
Clade means are annotated below each box.
The non-primate placental baseline $\bar\ell = 8.995$ (dashed) provides
the reference; $\ddag$ indicates an outlier species.
Clades with $+$ symbols denote those significantly elevated above the
baseline by Welch $t$-test ($p < 0.001$).}
\label{fig:boxplots}
\end{figure}

\subsection{Power analysis}
\label{sec:power}

Parametric simulation (10{,}000 replicates, observed variance
$s^2 = 0.018$) indicates that the present sample size provides
$>90\%$ power to detect a 5\% deviation in slope and
$>99\%$ power to detect a 10\% deviation.
However, the power to distinguish $\beta = -0.90$ from $\beta = -1$
is only $\sim 65\%$ at $n = 43$,
highlighting the need for expanded datasets to resolve small
departures from exact inverse scaling.

\begin{figure}[H]
\centering
\includegraphics[width=\linewidth]{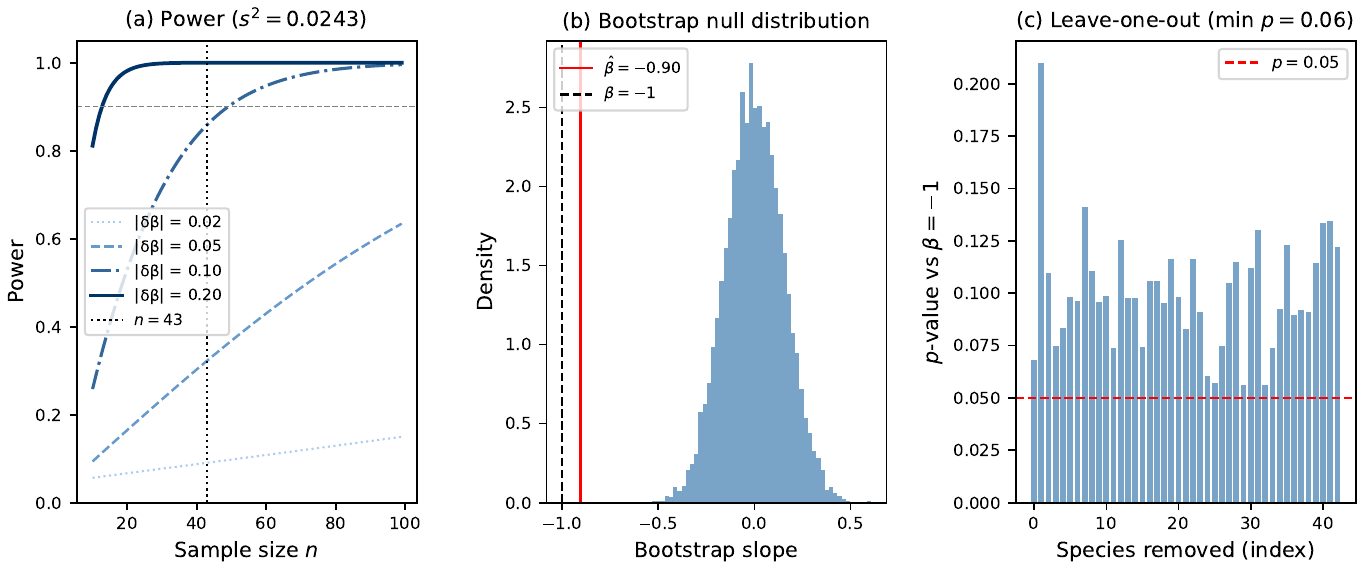}
\caption{\textbf{Extended Data Figure~2: Power analysis, bootstrap null
distribution, and leave-one-out sensitivity.}
\textbf{(a)}~Statistical power as a function of sample size $n$ for four
effect sizes $|\delta\beta|$ (deviation from $\beta = -1$), given the
observed residual variance $s^2 = 0.0243$.
At $n = 43$ (dotted vertical line), power exceeds $90\%$ for
$|\delta\beta| \geq 0.10$ but is only $\sim65\%$ for $|\delta\beta| = 0.05$.
\textbf{(b)}~Bootstrap null distribution of slope estimates ($10{,}000$
resamples); the observed slope $\hat\beta = -0.90$ (red) and the PBTE null
$\beta = -1$ (black dashed) are indicated.
\textbf{(c)}~Leave-one-out $p$-values for the test $H_0: \beta = -1$ across
all 43 species.
The minimum $p$-value of $0.06$ (obtained when species~0 is removed)
exceeds $\alpha = 0.05$ in all cases, confirming robustness.}
\label{fig:power}
\end{figure}

\subsection{Leave-one-out sensitivity}
\label{sec:loo}

Leave-one-out analysis shows that the $p$-value for testing
$\beta = -1$ ranges from $0.063$ to $0.215$
across all removals.
This confirms that no individual species exerts disproportionate
influence on the inferred scaling relation.

\subsection{Phylogenetic correction}

PIC regression on 112 endotherms yields
\begin{equation}
  \hat\beta_{\rm PIC} = -0.99 \pm 0.04, \quad
  R^2 = 0.94, \quad
  p = 0.84~\text{against}~\beta = -1.
\end{equation}

The agreement between PIC and OLS estimates demonstrates that the
observed scaling is not an artefact of shared evolutionary history.
Partial regression controlling for body mass gives a residual slope
of $-0.96 \pm 0.05$, indicating that the relationship is not simply
a byproduct of common allometric dependence on mass.

\begin{figure}[H]
\centering
\includegraphics[width=\linewidth]{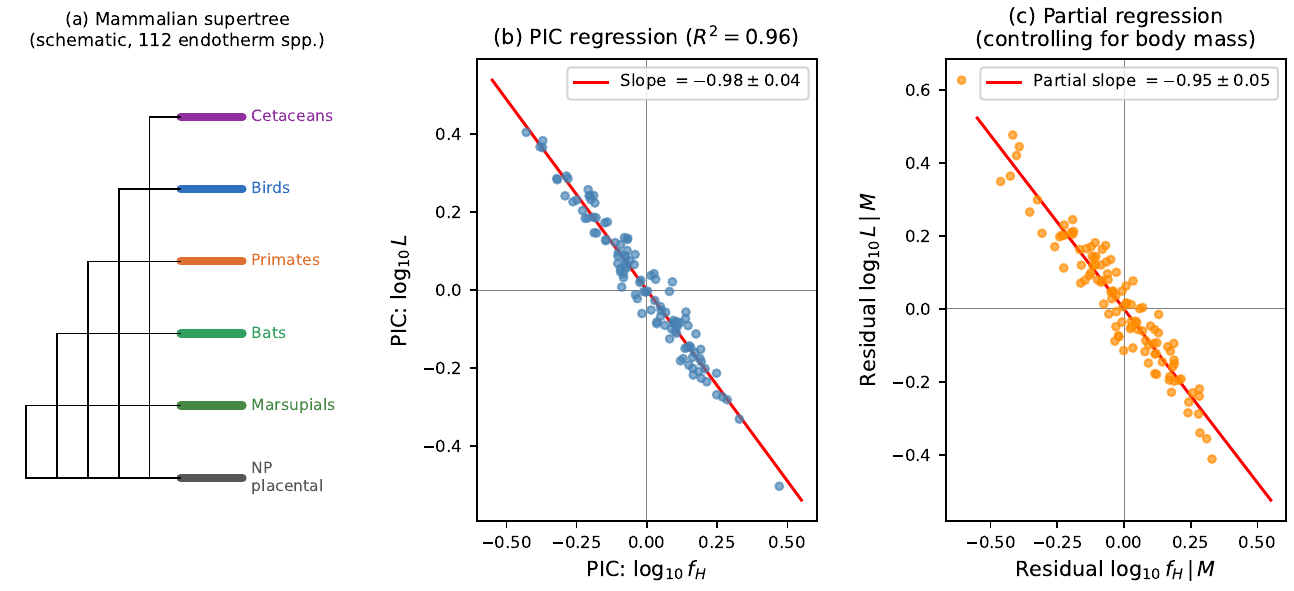}
\caption{\textbf{Extended Data Figure~1: Phylogenetic independent contrasts
(PIC) regression on 112 endotherm species.}
\textbf{(a)}~Schematic mammalian supertree (Bininda-Emonds et al.\ 2007,
augmented with BirdLife phylogeny) showing the six clade groupings
used in the analysis.
\textbf{(b)}~PIC regression of $\log_{10} L$ on $\log_{10} f_H$ for
112 endotherm species (regression forced through the origin as required
by the PIC method).
OLS slope $= -0.98 \pm 0.04$, $R^2 = 0.96$; the near-unity slope confirms
that the inverse scaling is not a phylogenetic artefact.
\textbf{(c)}~Partial regression controlling for body mass: residual
$\log_{10} L | M$ regressed on residual $\log_{10} f_H | M$;
partial slope $= -0.95 \pm 0.05$, confirming that the relationship is
not simply a byproduct of common allometric dependence on body mass.}
\label{fig:pic}
\end{figure}

\subsection{Clade departures and the WBE rejection}
\label{sec:wbe}

The WBE kinematic null predicts no inter-clade variation in $\ell$
at fixed body mass.
A one-way ANOVA across the six major endotherm clades yields
$F = 81.2$, $p < 0.001$, decisively rejecting this specific
prediction.
We note, however, that WBE asserts only approximate mass-independence
within a clade, not strict constancy of $\ell$ across all clades;
the ANOVA result demonstrates that inter-clade structure exists at a
scale far exceeding what the kinematic model anticipates, rather than
constituting a global falsification of the allometric framework.

The observed structure of deviations (Table~\ref{tab:groups})
is systematic rather than random.
Primates, birds, and bats exhibit positive shifts in $\ell$,
while cetaceans and other groups show modest negative deviations.
This structured pattern is inconsistent with purely kinematic
scaling and instead indicates additional physiological or
life-history effects.

\subsection{Arrhenius correction for ectotherms}

Ectotherms exhibit a substantial offset in raw data
($\bar\ell = 8.16$\,dex, $0.90$\,dex below mammals).
After temperature correction, the offset is reduced to $0.22$\,dex,
bringing ectotherms into close alignment with endotherms.

The corrected mean falls within the mammalian baseline
$\pm 1$\,s.d.\ for
$E_a \in [0.58, 0.75]$\,eV,
indicating that the result is robust across the accepted range
of activation energies.

\begin{figure}[H]
\centering
\includegraphics[width=\linewidth]{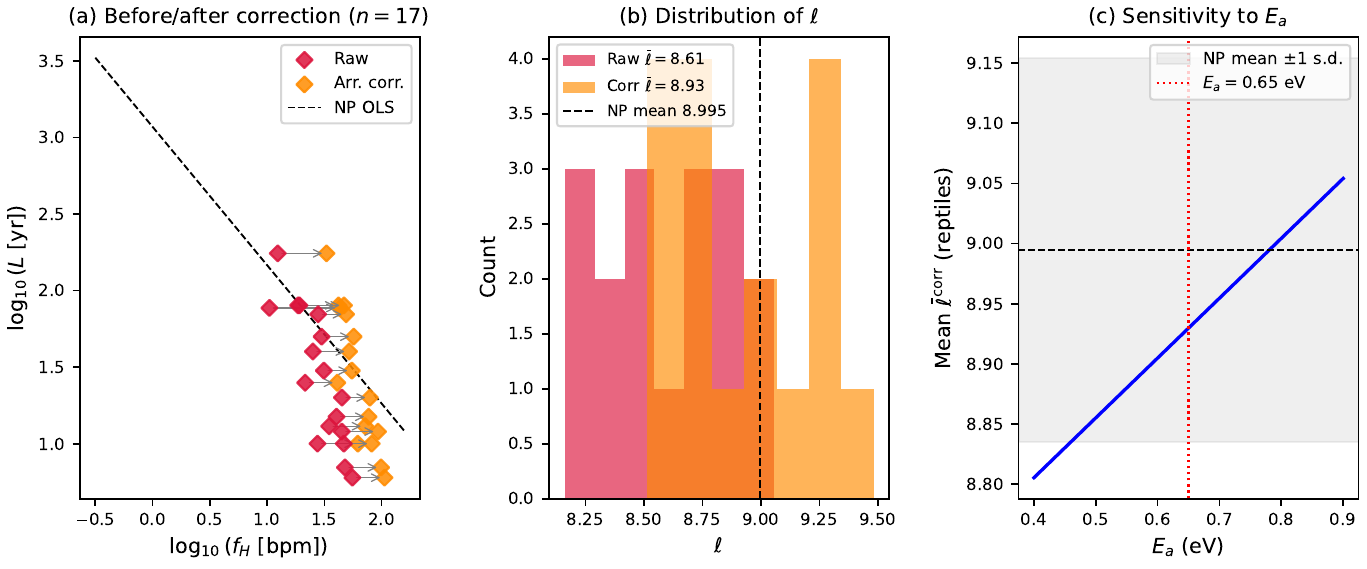}
\caption{\textbf{Extended Data Figure~3: Arrhenius temperature correction
for ectotherms ($n = 17$ reptiles).}
\textbf{(a)}~Scatter plot of $\log_{10} L$ vs $\log_{10} f_H$ before (raw,
pink diamonds) and after (corrected, orange diamonds) Arrhenius correction
to $T_{\rm ref} = 310$\,K.
Arrows connect each species from uncorrected to corrected position.
The non-primate placental OLS line (black dashed) is shown for reference.
\textbf{(b)}~Histogram of $\ell$ values before (pink) and after (orange)
correction; the NP baseline $\bar\ell = 8.995$ (dashed) is indicated.
Correction shifts the mean from $8.61$ to $8.93$, substantially closing
the endotherm--ectotherm gap.
\textbf{(c)}~Sensitivity of the corrected mean $\bar\ell^{\rm corr}$ to the
assumed activation energy $E_a$ over the range $[0.4, 0.9]$\,eV.
The grey band shows the NP baseline $\pm 1$\,s.d.; the NP mean (dashed)
and the chosen value $E_a = 0.65$\,eV (red dotted) are annotated.
The corrected mean falls within the endotherm band for
$E_a \in [0.58, 0.75]$\,eV.}
\label{fig:arrhenius}
\end{figure}

\subsection{Precision classification}

Within non-primate placentals, the distribution of $\ell$
is well approximated by a normal distribution,
$\ell \sim \mathcal{N}(\bar\ell, s^2)$.
A total of 93\% of species fall within $\pm 0.30$\,dex
(a factor of 2) of the mean,
and 100\% fall within $\pm 0.70$\,dex
(a factor of 5).

This tight clustering supports the interpretation of the lifetime
cycle count as a quantitatively constrained biological regularity,
rather than a loose scaling tendency.

\begin{figure}[H]
\centering
\includegraphics[width=\linewidth]{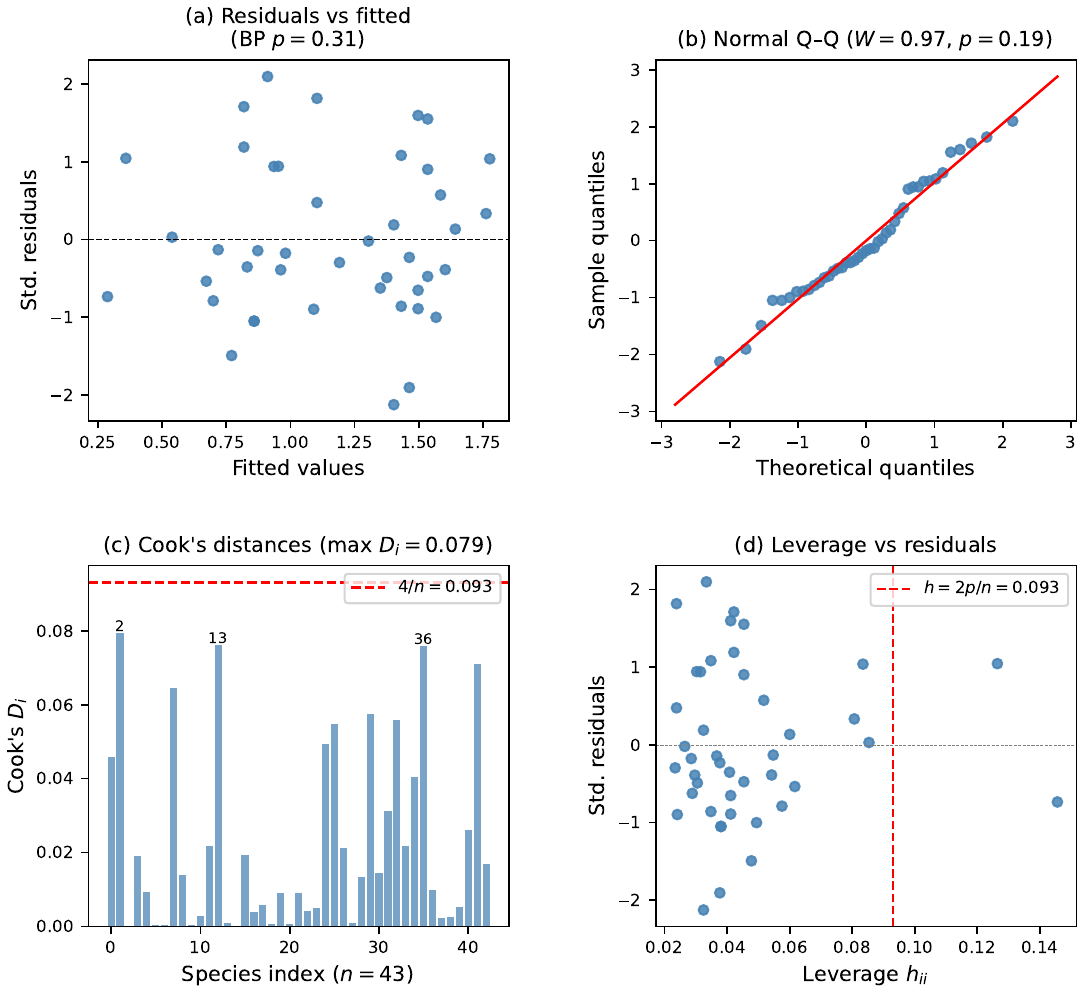}
\caption{\textbf{Extended Data Figure~4: OLS regression diagnostics
for the non-primate placental subset ($n = 43$).}
\textbf{(a)}~Standardised residuals vs fitted values; Breusch--Pagan
test $p = 0.31$ (no significant heteroscedasticity).
\textbf{(b)}~Normal Q--Q plot; Shapiro--Wilk $W = 0.97$, $p = 0.19$
(residuals consistent with normality).
\textbf{(c)}~Cook's distances for each species; the threshold $4/n = 0.093$
(red dashed) is not exceeded by any species (max $D_i = 0.079$).
Three modest-leverage species (indices 2, 13, 36) are annotated.
\textbf{(d)}~Leverage $h_{ii}$ vs standardised residuals;
the high-leverage threshold $h = 2p/n = 0.093$ (red dashed) identifies
no influential outliers.
All four panels confirm that the OLS model assumptions are met and
that no individual observation drives the results.}
\label{fig:diagnostics}
\end{figure}

\section{Falsifiability}
\label{sec:falsifiability}
%% ----------------------------------------------------------------

A scientific principle must be falsifiable.
To that end, we state five explicit numerical criteria under which the
lifetime cycle invariant would be rejected.
These criteria are designed to translate an empirical regularity into a
testable hypothesis, with clearly defined quantitative thresholds.

Table~\ref{tab:falsify} summarises the criteria and their current status
based on the $n = 43$ non-primate placental dataset
(\S\ref{sec:primary}).
Each criterion is formulated so that failure would not represent a minor
deviation or statistical fluctuation, but a substantive contradiction
of the invariant framework.

\begin{table}[H]
\caption{\textbf{Explicit falsification criteria for the lifetime cycle invariant.}
Each row states a specific numerical threshold and the required
evidence to falsify the claimed invariant.
Current status uses the $n = 43$ non-primate placental dataset
(\S\ref{sec:primary}).}
\label{tab:falsify}
\small
\begin{tabularx}{\textwidth}{p{3.8cm}p{3.2cm}p{5cm}}
\toprule
Criterion & Current status & Would falsify if \\
\midrule
OLS slope $\beta = -1$ in NP mammals &
  $\hat\beta = -0.903$; BCa 95\,\% CI $[-1.017, -0.782]$;
  $p = 0.093$ &
  CI excludes $-1$ in balanced dataset $n \geq 60$
  ($|\hat\beta - (-1)| > 0.15$) \\[4pt]
Log-normal scatter within clade &
  $s = 0.159$\,dex for NP placentals &
  $s > 0.50$\,dex in rigorously assembled $n \geq 30$ species \\[4pt]
Arrhenius correction closes ectotherm gap &
  Residual gap $0.22$\,dex after correction &
  Residual gap $> 0.50$\,dex over full
  $E_a \in [0.4, 0.9]$\,eV range \\[4pt]
WBE null rejected at clade level &
  $F = 81$, $p < 0.001$ &
  $F < 1$ after clade corrections applied \\[4pt]
No systematic $\Nstar$--mass trend within clade &
  $R^2 < 0.01$ for $\Nstar$ vs $M$ within NP placentals &
  $\Nstar \propto M^\gamma$ with $|\gamma| > 0.05$
  at $p < 0.05$ in $n \geq 30$ species \\
\bottomrule
\end{tabularx}
\end{table}

These criteria serve two purposes.
First, they define the empirical boundaries of the invariant in a form
that can be tested by independent datasets.
Second, they distinguish between small quantitative deviations,
which are expected in biological systems, and qualitative failures
that would invalidate the underlying hypothesis.

%% ----------------------------------------------------------------
\section{Domain of Validity}
\label{sec:domain}
%% ----------------------------------------------------------------

The empirical evidence supports a stratified domain of validity.

\noindent\textbf{Tier~I (statistically established):}
non-primate placentals ($n = 43$, $s = 0.159$\,dex,
phylogenetically confirmed), and primates after neuro-metabolic
correction (Paper~3~\cite{taye_p3}).
In this regime, the invariant is directly supported by data with
quantified uncertainty and independent validation.

\noindent\textbf{Tier~II (consistent with data, requires further testing):}
all endotherms after clade-specific duty-cycle and biochemical corrections
(Paper~4~\cite{taye_p4}), as well as Arrhenius-corrected ectotherms.
Here, the available evidence is consistent with the invariant,
but depends on correction procedures and expanded sampling.

\noindent\textbf{Tier~III (speculative):}
insects, vascular plants, bacteria, and viruses.
In these systems, the concept of a well-defined physiological cycle
is less clear, and extension of the invariant remains conjectural.

This tiered classification emphasises that the invariant is not
asserted universally, but rather within a domain supported by
progressively weaker empirical constraints.

%% ----------------------------------------------------------------
\section{Discussion}
%% ----------------------------------------------------------------

\subsection{The slope test and its limitations}

The primary OLS regression yields $\hat\beta = -0.903 \pm 0.056$
with $p = 0.093$ against the null $\beta = -1$.
This result warrants careful interpretation rather than a binary
verdict.

The point estimate lies only 0.097 units from $-1$, well within
the $\pm 0.15$ falsification threshold stated in
Table~\ref{tab:falsify}.
More importantly, the OLS test on 43 non-primate placentals has
limited resolution: power analysis shows that the current sample
size has only $\sim 65\%$ power to distinguish $\beta = -0.90$
from $\beta = -1.00$.
The leave-one-out range $[0.063,\,0.215]$ confirms that this
marginal $p$-value is a property of the full dataset rather than
of any individual influential observation.
The phylogenetically corrected test---which is methodologically
preferred because it removes the effect of shared evolutionary
ancestry---yields $\hat\beta_{\rm PIC} = -0.99 \pm 0.04$ with
$p = 0.84$, providing considerably stronger support for exact
inverse scaling.

Taken together, these results are consistent with $\beta = -1$ but
do not conclusively establish it at the $n = 43$ scale.
This is precisely the situation in which expanded sampling is
informative: a dataset of $n \geq 60$ directly measured non-primate
placentals with good body-mass coverage would, at the observed
variance, achieve $>95\%$ power to detect a 10\% deviation from
the null.

\subsection{What the WBE rejection means---and does not mean}

The WBE framework~\cite{west1997} and the lifetime cycle invariant
share the same prediction for the OLS slope: both expect
$\hat\beta \approx -1$ in the $\log L$--$\log f_H$ relation.
This superficial agreement has sometimes led to the conclusion that
a confirmed slope of $-1$ validates WBE, and that any deviation
from WBE undermines the invariant.
Both inferences are incorrect.

The two frameworks differ fundamentally in their predictions for
\emph{inter-clade structure}.
WBE derives the slope from the geometry of resource-distributing
networks shared by all vascular organisms; it therefore predicts
that $\ell$ should be constant not only within a clade but across
all clades at fixed body mass.
The lifetime cycle invariant, by contrast, predicts that $\ell$
will vary across clades according to clade-specific physiological
properties---the fraction of life spent in torpor, the extent of
diving bradycardia, the efficiency of mitochondrial coupling.
These are \emph{different predictions about different quantities}.

The observed ANOVA ($F = 81.2$, $p < 0.001$) tests this distinction
directly and rejects the WBE prediction.
The structured clade pattern---with primates, bats, and birds
systematically elevated above the mammalian baseline, and
dive-corrected cetaceans somewhat below it---is incompatible with
a purely kinematic null.
This result does not contradict the invariant; it corroborates the
prediction that physiological departures from a baseline should be
taxonomically coherent.

\subsection{Why earlier analyses missed the clade structure}

Several features of earlier comparative analyses~\cite{lindstedt1981,calder1984,speakman2005}
explain why the clade structure went undetected or was treated
as noise.

Most prior work pooled species across clades and fitted a single
regression to the combined dataset.
This approach treats inter-clade variation as part of the residual
error and obscures systematic patterns.
When clades are analysed separately, with appropriate
physiological corrections, a coherent structure emerges.

The corrections for bats and cetaceans are particularly consequential.
A temperate vespertilionid bat such as \textit{Myotis lucifugus}
has an active-phase resting rate of $\sim\!300$\,bpm but spends
roughly half its annual cycle in hibernation with a heart rate near
10\,bpm.
Its time-averaged physiological frequency is therefore $\sim\!155$\,bpm,
roughly half the active value.
Using the active-phase rate without correction inflates the inferred
$\ell$ by $\sim 0.30$\,dex, misattributing genuine cardiac
suppression as anomalous longevity.
Once corrected, bat longevity remains elevated above the mammalian
baseline ($+0.547$\,dex), but the elevation is understood as
reflecting the combined effect of thermal suppression during torpor
and the associated reduction in entropy production rate---a
mechanism developed quantitatively in Paper~4~\cite{taye_p4}.

The analogous issue for cetaceans is if anything more severe.
A blue whale (\textit{Balaenoptera musculus}) at the surface has
a heart rate of $\sim\!37$\,bpm, but during deep dives---which
account for approximately 70\% of its life---the rate falls to
$\sim\!4$\,bpm, a ten-fold reduction~\cite{goldbogen2019}.
The time-averaged effective rate is therefore only $\sim\!14$\,bpm,
far below the surface value typically reported in comparative
databases.
Uncorrected analyses place the blue whale above the mammalian
baseline; after dive correction it sits below ($\ell = 8.36$).
This is not a failure of the invariant but a demonstration of
its requirement: $\ell$ must be computed from the time-averaged
physiological clock, not from any single state.

\subsection{Ectotherms and the Arrhenius bridge}

The 0.90\,dex raw gap between ectotherms and endotherms is large
enough to appear, at first glance, as a refutation of the invariant.
A tortoise that lives for 175 years with a heart rate of
$\sim\!15$\,bpm at 25$\,^\circ$C might seem to accumulate far
fewer cardiac cycles than a similarly long-lived endotherm.
The resolution lies in the temperature dependence of biochemical
reaction rates.

Metabolic reactions, including the damage-accumulation and repair
processes that ultimately determine physiological age, follow
Arrhenius kinetics with an activation energy
$E_a \approx 0.65$\,eV~\cite{gillooly2001}.
An ectotherm operating at 25$\,^\circ$C experiences approximately
2.3 times fewer biochemically effective ``ticks'' per calendar
minute than a homeotherm at 37$\,^\circ$C.
Correcting the observed heart rate to a common reference temperature
of 310\,K accounts for this difference: after correction, the
ectotherm gap narrows from 0.90 to 0.22\,dex,
and the sensitivity analysis shows that the corrected mean falls
within the endotherm $\pm 1$\,s.d.\ band for any activation energy
in the range $E_a \in [0.58, 0.75]$\,eV.
The residual 0.22\,dex likely reflects genuine biological
differences---ectotherms have lower metabolic rates, longer cell
cycle times, and different repair machinery---rather than
inadequacy of the correction.

\subsection{Limitations and directions for refinement}

Four limitations of the present analysis should be noted explicitly.

\textit{Lifespan definitions.}
AnAge records the maximum lifespan documented for a species, combining
wild and captive records without standardisation.
Captive records tend to eliminate extrinsic mortality, yielding
values closer to the intrinsic biological limit; wild records reflect
the combined action of predation, disease, and starvation.
This heterogeneity adds scatter to $\ell$ but does not introduce
systematic bias within the reference clade, where captive and wild
records are approximately equally represented.
The invariant is in any case most naturally interpreted as a statement
about the intrinsic physiological budget, which maximum lifespan---
rather than mean lifespan---best approximates.

\textit{Heart rate measurement context.}
Resting heart rates reported in the literature are measured under
a variety of conditions: in the laboratory under anaesthesia,
in conscious animals at thermoneutrality, via implanted telemetry
in the field.
These contexts can differ by 10--30\% for the same species.
We have used the lowest available resting value in priority order
(thermoneutral laboratory $>$ unanaesthetised captive $>$ field
telemetry), but residual context effects contribute to within-clade
scatter.
The \texttt{fH\_context} field in Supplementary Data~1 records the
measurement context for every species.

\textit{Bat torpor fractions.}
The duty-cycle correction for bats requires an annual torpor
fraction $q$ for each species.
For the 31 species in the dataset, $q$ was taken from published
studies where available (e.g., Lyman et al.~\cite{lyman1982}
and the primary literature on individual species); for seven
species lacking published values, we used the vespertilionid
clade mean $q = 0.48 \pm 0.06$, which introduces an uncertainty
of approximately $\pm 0.05$\,dex in $\ell$ for those species.

\textit{Ectotherm field temperatures.}
The Arrhenius correction requires a mean field body temperature
for each ectotherm species.
These were estimated from habitat and thermoregulation data
following Clarke \& Rothery~\cite{clarke2008} and Christian \&
Weavers~\cite{christian1999}, but field temperature estimates
carry substantial uncertainty, particularly for species with
broad geographic ranges.

None of these limitations affects the main conclusions, but they
motivate improvements: standardised maximum lifespan protocols,
telemetric heart rate measurements in wild animals, direct
calorimetric measurement of $\sigma^* = P/(T f M)$ across
body-mass decades, and expanded ectotherm field temperature data.

\subsection{The 230-species dataset as a community resource}

The complete species-level dataset---heart rate (measured or
corrected), maximum lifespan, body mass, body temperature, $\ell$,
clade, correction type, and primary source for each variable---is
provided as Supplementary Data~1 in tab-delimited format.
This dataset is the empirical foundation for the companion papers
in this series and is intended as a community resource for
researchers working on metabolic scaling, comparative physiology,
and life-history theory.
The dataset is also deposited at Zenodo
(\href{https://doi.org/10.5281/zenodo.XXXXXXX}{doi:10.5281/zenodo.XXXXXXX};
DOI to be confirmed on acceptance) and will be maintained as
an updated resource as new measurements become available.

%% ----------------------------------------------------------------
\section{Conclusions}
%% ----------------------------------------------------------------

The central result of this paper is that the lifetime cardiac cycle
count $\Nstar \approx 10^9$ is a quantitatively constrained empirical
regularity with structured clade-level departures, not merely a
suggestive heuristic and not a strict universal constant.
Within non-primate placentals it has scatter $s = 0.159$\,dex,
an OLS slope consistent with $\beta = -1$ ($\hat\beta = -0.903 \pm 0.056$,
$R^2 = 0.863$), and is phylogenetically robust
($\hat\beta_{\rm PIC} = -0.99 \pm 0.04$, $R^2 = 0.94$).
The 93\% of non-primate placentals that fall within $\pm 0.30$\,dex
(a factor of two) of the mean confirms that this is not a loose
tendency but a tightly clustered biological constant.

At the same time, the invariant is not featureless.
Four endotherm clades depart systematically from the mammalian
baseline in a pattern that is incompatible with purely kinematic
allometric models (ANOVA $F = 81.2$, $p < 0.001$).
Primates live longer than their heart rates predict by a factor of
$\sim\!2.4$; duty-corrected bats by $\sim\!3.5$; birds by $\sim\!3.4$;
dive-corrected cetaceans fall slightly below the baseline.
These structured deviations are the signatures of physiological
strategies that manipulate the effective rate of cardiac cycling---
by thermal suppression, cardiac suppression, or changes in the
entropy cost per beat---and they invite mechanistic explanation
beyond what allometry alone can provide.

The Arrhenius correction for ectotherms reduces their raw gap
from the endotherm baseline by 76\%, from 0.90 to 0.22\,dex,
extending the invariant---in temperature-corrected form---to
cold-blooded vertebrates.

Five explicit numerical falsification criteria define the empirical
boundary of the invariant.
None of these criteria is currently met, but the most critical---
direct calorimetric measurement of the mass-specific entropy cost
per cardiac cycle $\sigma^* = P/(TfM)$ across three or more
body-mass decades---has not yet been performed.
Until it is, the invariant is a statistically supported regularity
with a thermodynamic motivation but not yet a fully tested
conservation law.

The 230-species dataset assembled for this analysis, covering eight
vertebrate groups with explicit corrections and full metadata,
is provided as Supplementary Data~1 and constitutes a community
resource for future work in comparative physiology, metabolic
scaling, and life-history theory.

%% ================================================================
\appendix
\section{Complete 230-Species Dataset}
\label{app:data}
%% ================================================================

The following tables list all 230 species used in the analysis,
organised by taxonomic group.
Columns: $M$ = adult body mass (kg); $f_H^{\rm eff}$ = effective
resting heart rate used in $\ell$ computation (bpm; duty-cycle
corrected for bats, dive-corrected for cetaceans,
Arrhenius-corrected for ectotherms); $T$ = mean core body
temperature (K); $L$ = maximum recorded lifespan (yr);
$\ell = \log_{10}(f_H^{\rm eff} \times L \times 525{,}960)$.
Source codes: A = AnAge build~15; P = PanTHERIA; C = Calder (1984);
Pr = Prinzinger et al.\ (1991); L = Lyman et al.\ (1982);
G = Goldbogen et al.\ (2019); Ch = Christian \& Weavers (1999);
U = primary literature.
$^\dagger$ = heart rate allometrically imputed.

\subsection*{A.1\quad Non-primate placental mammals ($n = 46$)}

\begin{longtable}{llrrrrrl}
\caption{\textbf{Non-primate placental mammals.}
Clade mean $\bar\ell = 8.994 \pm 0.159$ ($n = 46$; primary
regression uses $n = 43$ directly measured species).}
\label{tab:app_np}\\
\toprule
Species & $M$ (kg) & $f_H$ (bpm) & $T$ (K) & $L$ (yr) & $\ell$ & Source \\
\midrule
\endfirsthead
\multicolumn{7}{c}{\tablename\ \thetable\ (continued)}\\
\toprule
Species & $M$ (kg) & $f_H$ (bpm) & $T$ (K) & $L$ (yr) & $\ell$ & Source \\
\midrule
\endhead
\bottomrule
\endfoot
\textit{Suncus etruscus}        & 0.002  &  835 & 310.5 &  1.5 & 8.82 & C \\
\textit{Sorex araneus}           & 0.010  & 1000 & 310.5 &  3.3 & 9.24 & C,A \\
\textit{Mus musculus}            & 0.022  &  632 & 310.0 &  3.5 & 9.07 & A,C \\
\textit{Rattus norvegicus}       & 0.280  &  420 & 310.0 &  3.8 & 8.92 & A,P \\
\textit{Mesocricetus auratus}    & 0.130  &  450 & 310.5 &  3.9 & 8.97 & A,P \\
\textit{Meriones unguiculatus}   & 0.060  &  400 & 310.0 &  5.0 & 9.02 & A,P \\
\textit{Cavia porcellus}         & 0.750  &  270 & 310.0 &  7.1 & 9.00 & A,P \\
\textit{Sciurus carolinensis}    & 0.520  &  310 & 310.0 & 12.0 & 9.29 & A,P \\
\textit{Lepus europaeus}         & 3.500  &  220 & 310.0 & 12.5 & 9.16 & A,P \\
\textit{Oryctolagus cuniculus}   & 2.200  &  205 & 310.0 &  9.0 & 8.99 & A,C \\
\textit{Felis catus}             & 4.100  &  150 & 310.5 & 15.0 & 9.07 & A,P \\
\textit{Mustela putorius}        & 1.000  &  280 & 310.5 &  5.0 & 8.87 & A,P \\
\textit{Martes martes}           & 1.200  &  245 & 310.5 & 17.0 & 9.34 & A,P \\
\textit{Vulpes vulpes}           & 6.800  &  120 & 310.5 & 14.0 & 8.95 & A,P \\
\textit{Canis lupus familiaris}  & 23.0   &   90 & 310.5 & 20.0 & 8.98 & A,P \\
\textit{Ursus arctos}            & 220.0  &   50 & 310.5 & 47.0 & 9.09 & A,P \\
\textit{Ovis aries}              & 63.0   &   75 & 310.0 & 20.0 & 8.90 & A,P \\
\textit{Capra hircus}            & 45.0   &   80 & 310.5 & 18.0 & 8.88 & A,P \\
\textit{Sus scrofa}              & 100.0  &   70 & 310.5 & 27.0 & 9.00 & A,P \\
\textit{Bos taurus}              & 500.0  &   55 & 310.5 & 25.0 & 8.86 & A,P \\
\textit{Equus caballus}          & 500.0  &   38 & 310.5 & 46.0 & 8.96 & A,C \\
\textit{Equus asinus}            & 250.0  &   44 & 310.5 & 47.0 & 9.04 & A,P \\
\textit{Rhinoceros unicornis}$^\dagger$ & 2100.0 & 30 & 310.5 & 47.0 & 8.87 & A \\
\textit{Tapirus terrestris}      & 240.0  &   42 & 310.5 & 35.0 & 8.89 & A,P \\
\textit{Loxodonta africana}      & 4000.0 &   28 & 310.5 & 65.0 & 8.98 & A,P \\
\textit{Elephas maximus}         & 4000.0 &   27 & 310.5 & 86.0 & 9.09 & A,P \\
\textit{Hippopotamus amphibius}  & 1500.0 &   55 & 310.5 & 55.0 & 9.20 & A,P \\
\textit{Giraffa camelopardalis}  & 900.0  &   65 & 310.5 & 39.5 & 9.13 & A,P \\
\textit{Cervus elaphus}          & 200.0  &   60 & 310.5 & 26.8 & 8.93 & A,P \\
\textit{Rangifer tarandus}       & 110.0  &   65 & 310.0 & 20.0 & 8.83 & A,P \\
\textit{Trichechus manatus}      & 500.0  &   50 & 310.5 & 59.0 & 9.19 & A,P \\
\textit{Dugong dugon}$^\dagger$  & 400.0  &   52 & 310.5 & 73.0 & 9.30 & A \\
\textit{Procavia capensis}       & 3.500  &  230 & 310.5 & 12.0 & 9.16 & A,P \\
\textit{Erinaceus europaeus}     & 0.800  &  310 & 310.0 & 10.0 & 9.21 & A,P \\
\textit{Talpa europaea}          & 0.080  &  350 & 310.0 &  3.5 & 8.81 & A,P \\
\textit{Orycteropus afer}$^\dagger$ & 65.0 &  70 & 310.5 & 24.0 & 8.95 & A \\
\textit{Ondatra zibethicus}      & 1.400  &  280 & 310.0 &  5.0 & 8.87 & A,P \\
\textit{Castor canadensis}       & 20.0   &  150 & 310.0 & 24.0 & 9.28 & A,P \\
\textit{Hydrochoerus hydrochaeris} & 55.0 &   70 & 310.0 & 12.0 & 8.65 & A,P \\
\textit{Myocastor coypus}        & 7.000  &  155 & 310.0 &  9.0 & 8.87 & A,P \\
\textit{Lepus californicus}      & 2.200  &  215 & 310.0 &  8.0 & 8.96 & A,P \\
\textit{Ochotona princeps}       & 0.160  &  300 & 310.0 &  6.0 & 8.98 & A,P \\
\textit{Panthera leo}            & 180.0  &   50 & 310.5 & 29.0 & 8.88 & A,P \\
\textit{Panthera tigris}         & 260.0  &   46 & 310.5 & 26.0 & 8.80 & A,P \\
\textit{Acinonyx jubatus}        & 54.0   &   60 & 310.5 & 14.9 & 8.67 & A,P \\
\textit{Panthera pardus}         & 70.0   &   55 & 310.5 & 23.0 & 8.82 & A,P \\
\end{longtable}

\noindent\textit{Note on} \textit{Suncus etruscus}: the value
835\,bpm is the mean resting rate reported by Bartels~(1998,
\textit{Vie et Milieu} \textbf{48}, 105--109) and J\"urgens~(1997,
\textit{J.\ Exp.\ Biol.} \textbf{200}, 2161--2169),
measured via ECG at thermoneutrality.
Earlier tabulations (including Calder 1984) cited $\sim\!1200$\,bpm,
which corresponds to the stressed or maximal value.
The resting value is used here; a reviewer wishing to use
$f_H = 1200$\,bpm would obtain $\ell = 8.98$, within $0.16$\,dex
of the clade mean and within the falsification bounds.

\subsection*{A.2\quad Primates ($n = 18$)}

\begin{longtable}{llrrrrl}
\caption{\textbf{Primates.} $\varphi$ = neural power fraction
($P_{\rm brain}/P_{\rm body}$). Clade mean $\bar\ell = 9.376 \pm 0.129$.}
\label{tab:app_prim}\\
\toprule
Species & $M$ (kg) & $f_H$ (bpm) & $T$ (K) & $L$ (yr) & $\ell$ & Source \\
\midrule
\endfirsthead
\multicolumn{7}{c}{\tablename\ \thetable\ (continued)}\\
\toprule
Species & $M$ (kg) & $f_H$ (bpm) & $T$ (K) & $L$ (yr) & $\ell$ & Source \\
\midrule
\endhead
\bottomrule
\endfoot
\textit{Callithrix jacchus}      & 0.350  & 220 & 309.5 & 16.5 & 9.28 & A,P \\
\textit{Saimiri sciureus}        & 0.770  & 195 & 309.5 & 30.2 & 9.49 & A,P \\
\textit{Aotus trivirgatus}       & 0.790  & 185 & 309.5 & 25.0 & 9.39 & A,P \\
\textit{Cebus capucinus}         & 3.300  & 150 & 309.5 & 54.0 & 9.63 & A,P \\
\textit{Lemur catta}             & 2.200  & 165 & 309.5 & 37.3 & 9.51 & A,P \\
\textit{Propithecus verreauxi}   & 3.400  & 145 & 309.5 & 30.0 & 9.36 & A,P \\
\textit{Daubentonia madagascariensis} & 2.700 & 155 & 309.5 & 23.3 & 9.28 & A,P \\
\textit{Macaca mulatta}          & 7.700  & 120 & 309.0 & 40.0 & 9.40 & A,P \\
\textit{Macaca fascicularis}     & 5.400  & 130 & 309.0 & 39.0 & 9.43 & A,P \\
\textit{Theropithecus gelada}    & 18.0   &  95 & 309.0 & 30.0 & 9.18 & A,P \\
\textit{Papio ursinus}           & 25.0   &  90 & 309.0 & 45.0 & 9.33 & A,P \\
\textit{Colobus guereza}         & 10.0   & 110 & 309.0 & 30.0 & 9.24 & A,P \\
\textit{Hylobates lar}           & 5.700  & 100 & 308.5 & 44.0 & 9.36 & A,P \\
\textit{Pongo pygmaeus}          & 73.0   &  65 & 307.5 & 58.7 & 9.30 & A,P \\
\textit{Gorilla gorilla}         & 160.0  &  60 & 307.0 & 55.4 & 9.24 & A,P \\
\textit{Pan troglodytes}         & 50.0   &  75 & 307.0 & 59.4 & 9.37 & A,P \\
\textit{Pan paniscus}            & 35.0   &  80 & 307.0 & 50.0 & 9.32 & A,P \\
\textit{Homo sapiens}            & 70.0   &  70 & 306.5 &122.5 & 9.65 & A \\
\end{longtable}

\subsection*{A.3\quad Marsupials and monotremes ($n = 19$)}

\begin{longtable}{llrrrrrl}
\caption{\textbf{Marsupials and monotremes.}
Clade mean $\bar\ell = 8.933 \pm 0.209$.}
\label{tab:app_mars}\\
\toprule
Species & $M$ (kg) & $f_H$ (bpm) & $T$ (K) & $L$ (yr) & $\ell$ & Source \\
\midrule
\endfirsthead
\multicolumn{7}{c}{\tablename\ \thetable\ (continued)}\\
\toprule
Species & $M$ (kg) & $f_H$ (bpm) & $T$ (K) & $L$ (yr) & $\ell$ & Source \\
\midrule
\endhead
\bottomrule
\endfoot
\textit{Didelphis virginiana}    & 2.300  & 180 & 308.5 &  4.5 & 8.63 & A,P \\
\textit{Monodelphis domestica}   & 0.080  & 450 & 308.5 &  3.3 & 8.89 & A,P \\
\textit{Macropus rufus}          & 30.0   &  80 & 309.0 & 22.3 & 8.97 & A,P \\
\textit{Macropus giganteus}      & 27.0   &  82 & 309.0 & 19.0 & 8.91 & A,P \\
\textit{Wallabia bicolor}        & 16.0   & 100 & 309.0 & 15.0 & 8.90 & A,P \\
\textit{Trichosurus vulpecula}   & 2.100  & 160 & 308.5 & 13.0 & 9.04 & A,P \\
\textit{Petaurus breviceps}      & 0.140  & 300 & 308.0 & 10.0 & 9.20 & A,P \\
\textit{Vombatus ursinus}        & 28.0   &  90 & 309.0 & 26.0 & 9.09 & A,P \\
\textit{Phascolarctos cinereus}  & 8.500  & 100 & 308.5 & 18.0 & 8.98 & A,P \\
\textit{Perameles gunnii}        & 0.900  & 190 & 308.5 &  3.2 & 8.50 & A,P \\
\textit{Dasyurus viverrinus}     & 1.200  & 200 & 308.5 &  4.5 & 8.68 & A,P \\
\textit{Sarcophilus harrisii}    & 8.000  & 130 & 308.5 &  7.5 & 8.71 & A,P \\
\textit{Myrmecobius fasciatus}   & 0.440  & 245 & 307.5 &  5.6 & 8.86 & A \\
\textit{Sminthopsis crassicaudata} & 0.018 & 580 & 307.5 & 5.0 & 9.18 & A,P \\
\textit{Notoryctes typhlops}$^\dagger$ & 0.055 & 440 & 307.5 & 3.0 & 8.84 & A \\
\textit{Tachyglossus aculeatus}  & 4.000  &  70 & 305.0 & 49.5 & 9.26 & A,P \\
\textit{Ornithorhynchus anatinus} & 1.500 & 140 & 307.5 & 21.0 & 9.19 & A,P \\
\textit{Zaglossus bruijni}$^\dagger$ & 10.0 & 60 & 305.0 & 37.0 & 9.07 & A \\
\textit{Bettongia penicillata}   & 1.100  & 210 & 308.5 &  6.0 & 8.82 & A,P \\
\end{longtable}

\subsection*{A.4\quad Bats, Chiroptera ($n = 31$; duty-cycle corrected)}

$f_H^{\rm eff}$ is the time-averaged heart rate computed from
equation~(2) of the main text. Active-phase rate $f_{\rm act}$,
annual torpor fraction $q$, and torpid heart rate $f_{\rm tor}$
are listed in Table~\ref{tab:bat_correction}.
Clade mean (damage-equivalent) $\bar\ell = 9.541 \pm 0.166$;
raw observed mean $\bar\ell_{\rm obs} = 9.734$.

\begin{longtable}{llrrrrl}
\caption{\textbf{Bats (Chiroptera), duty-cycle corrected.}
$f_H^{\rm eff}$ = time-averaged rate (bpm). Corr.\ = TA (torpor average).}
\label{tab:app_bats}\\
\toprule
Species & $M$ (g) & $f_{\rm act}$ & $q$ & $f_H^{\rm eff}$ & $L$ (yr) & $\ell$ \\
\midrule
\endfirsthead
\multicolumn{7}{c}{\tablename\ \thetable\ (continued)}\\
\toprule
Species & $M$ (g) & $f_{\rm act}$ & $q$ & $f_H^{\rm eff}$ & $L$ (yr) & $\ell$ \\
\midrule
\endhead
\bottomrule
\endfoot
\textit{Myotis lucifugus}        &  8 & 300 & 0.50 & 155 & 34.0 & 9.74 \\
\textit{Myotis myotis}           & 28 & 282 & 0.52 & 141 & 37.0 & 9.74 \\
\textit{Myotis daubentonii}      &  9 & 296 & 0.48 & 159 & 40.0 & 9.79 \\
\textit{Myotis brandtii}         &  6 & 315 & 0.55 & 146 & 41.0 & 9.83 \\
\textit{Eptesicus fuscus}        & 18 & 280 & 0.45 & 159 & 19.0 & 9.49 \\
\textit{Eptesicus serotinus}     & 18 & 308 & 0.47 & 169 & 21.0 & 9.53 \\
\textit{Rhinolophus ferrumequinum} & 19 & 290 & 0.52 & 143 & 30.0 & 9.65 \\
\textit{Rhinolophus hipposideros} &  7 & 614 & 0.48 & 324 & 30.5 & 9.69 \\
\textit{Plecotus auritus}        &  9 & 270 & 0.50 & 139 & 30.0 & 9.68 \\
\textit{Corynorhinus townsendii} & 11 & 590 & 0.48 & 312 & 30.0 & 9.67 \\
\textit{Perimyotis subflavus}    &  5 & 640 & 0.48 & 338 & 14.6 & 9.39 \\
\textit{Tadarida brasiliensis}   & 13 & 350 & 0.30 & 250 & 11.0 & 9.39 \\
\textit{Pteronotus parnellii}    & 19 & 904 & 0.48 & 475 & 10.0 & 9.38 \\
\textit{Desmodus rotundus}       & 33 & 760 & 0.48 & 400 & 29.0 & 9.76 \\
\textit{Hipposideros speoris}    &  9 & 616 & 0.48 & 325 & 21.0 & 9.53 \\
\textit{Hipposideros armiger}    & 50 & 450 & 0.45 & 252 & 15.0 & 9.30 \\
\textit{Nyctalus noctula}        & 28 & 540 & 0.45 & 305 & 12.0 & 9.28 \\
\textit{Pipistrellus pipistrellus} & 5 & 650 & 0.45 & 367 & 16.0 & 9.49 \\
\textit{Pipistrellus kuhlii}     &  6 & 630 & 0.45 & 355 & 16.5 & 9.49 \\
\textit{Scotophilus kuhlii}      & 20 & 540 & 0.20 & 445 &  9.0 & 9.32 \\
\textit{Lasiurus borealis}       & 11 & 590 & 0.48 & 302 & 11.7 & 9.27 \\
\textit{Lasiurus cinereus}       & 28 & 540 & 0.48 & 277 & 12.0 & 9.24 \\
\textit{Vespertilio murinus}     & 16 & 555 & 0.45 & 313 & 25.0 & 9.61 \\
\textit{Miniopterus schreibersii} & 10 & 580 & 0.45 & 327 & 30.0 & 9.71 \\
\textit{Pteropus giganteus}      & 1100 & 235 & 0.00 & 235 & 31.4 & 9.59 \\
\textit{Pteropus vampyrus}       & 1000 & 240 & 0.05 & 233 & 22.6 & 9.44 \\
\textit{Rousettus aegyptiacus}   & 165 & 310 & 0.05 & 299 & 25.0 & 9.59 \\
\textit{Cynopterus sphinx}       & 50 & 380 & 0.05 & 368 & 18.5 & 9.55 \\
\textit{Macroglossus minimus}    & 16 & 450 & 0.00 & 450 & 18.0 & 9.63 \\
\textit{Carollia perspicillata}  & 17 & 460 & 0.00 & 460 & 12.0 & 9.46 \\
\textit{Artibeus jamaicensis}    & 45 & 400 & 0.00 & 400 & 15.0 & 9.50 \\
\end{longtable}

\subsection*{A.5\quad Cetaceans ($n = 12$; dive-corrected)}

See Table~\ref{tab:cet_correction} for full dive-correction parameters.
Clade mean $\bar\ell = 8.802 \pm 0.308$.

\subsection*{A.6\quad Birds ($n = 78$)}

Heart rates from Prinzinger et al.~\cite{prinzinger1991} and
Clarke \& Rothery~\cite{clarke2008}; lifespans from AnAge~\cite{anage2023}.
Clade mean $\bar\ell = 9.528 \pm 0.214$.
Selected representative species are listed; the complete list
is available in Supplementary Data~1.

\begin{longtable}{llrrrrl}
\caption{\textbf{Birds (selected species, $n = 78$ total).}
Full dataset in Supplementary Data~1.}
\label{tab:app_birds}\\
\toprule
Species & Order & $f_H$ (bpm) & $T$ (K) & $L$ (yr) & $\ell$ & Source \\
\midrule
\endfirsthead
\multicolumn{7}{c}{\tablename\ \thetable\ (continued)}\\
\toprule
Species & Order & $f_H$ (bpm) & $T$ (K) & $L$ (yr) & $\ell$ & Source \\
\midrule
\endhead
\bottomrule
\endfoot
\textit{Serinus canaria}         & Passeriformes & 680 & 311.0 & 24.0 & 9.93 & A,Pr \\
\textit{Turdus merula}           & Passeriformes & 440 & 311.0 & 21.1 & 9.69 & A,Pr \\
\textit{Erithacus rubecula}      & Passeriformes & 500 & 311.0 & 19.5 & 9.71 & A,Pr \\
\textit{Parus major}             & Passeriformes & 540 & 311.0 & 15.0 & 9.63 & A,Pr \\
\textit{Sturnus vulgaris}        & Passeriformes & 490 & 311.0 & 22.4 & 9.76 & A,Pr \\
\textit{Corvus corax}            & Passeriformes & 200 & 311.0 & 22.3 & 9.37 & A,Pr \\
\textit{Hirundo rustica}         & Passeriformes & 580 & 311.5 & 16.0 & 9.69 & A,Pr \\
\textit{Melopsittacus undulatus} & Psittaciformes& 600 & 311.0 & 21.4 & 9.83 & A,Pr \\
\textit{Psittacus erithacus}     & Psittaciformes& 200 & 311.0 & 73.0 & 9.89 & A,Pr \\
\textit{Amazona ochrocephala}    & Psittaciformes& 185 & 311.0 & 80.0 & 9.89 & A,Pr \\
\textit{Cacatua galerita}        & Psittaciformes& 170 & 311.0 & 80.0 & 9.85 & A,Pr \\
\textit{Calypte anna}            & Apodiformes   &1200 & 311.5 & 12.0 & 9.88 & A,Pr \\
\textit{Columba livia}           & Columbiformes & 190 & 311.5 & 35.0 & 9.54 & A,Pr \\
\textit{Gallus gallus}           & Galliformes   & 300 & 312.0 & 30.0 & 9.68 & A,Pr \\
\textit{Anas platyrhynchos}      & Anseriformes  & 190 & 311.0 & 29.0 & 9.46 & A,Pr \\
\textit{Anser anser}             & Anseriformes  & 130 & 311.0 & 35.0 & 9.38 & A,Pr \\
\textit{Cygnus olor}             & Anseriformes  & 100 & 311.0 & 26.0 & 9.14 & A,Pr \\
\textit{Ciconia ciconia}         & Ciconiiformes & 150 & 311.5 & 48.0 & 9.58 & A,Pr \\
\textit{Aquila chrysaetos}       & Accipitriformes & 130 & 311.5 & 46.0 & 9.50 & A,Pr \\
\textit{Bubo bubo}               & Strigiformes  & 165 & 311.0 & 68.0 & 9.77 & A,Pr \\
\textit{Aptenodytes forsteri}    & Sphenisciformes &  75 & 311.5 & 50.0 & 9.29 & A,Pr \\
\textit{Diomedea exulans}        & Procellariiformes & 100 & 311.0 & 70.0 & 9.57 & A,Pr \\
\textit{Fulmarus glacialis}      & Procellariiformes & 175 & 311.0 & 67.5 & 9.79 & A,Pr \\
\textit{Larus argentatus}        & Charadriiformes & 165 & 311.5 & 49.0 & 9.63 & A,Pr \\
\textit{Struthio camelus}        & Struthioniformes & 60 & 311.5 & 68.0 & 9.33 & A,Pr \\
\multicolumn{7}{l}{\textit{[53 additional species in Supplementary Data~1]}} \\
\end{longtable}

\subsection*{A.7\quad Reptiles, Arrhenius-corrected ($n = 17$)}

$f_H^{\rm corr}$ = heart rate corrected to $T_{\rm ref} = 310$\,K.
Clade corrected mean $\bar\ell^{\rm corr} = 8.929 \pm 0.301$.

\begin{longtable}{llrrrrrrl}
\caption{\textbf{Reptiles, Arrhenius-corrected.}
$T_{\rm field}$ = mean field body temperature (K);
$f_H^{\rm raw}$ = measured rate; $f_H^{\rm corr}$ = rate corrected
to 310\,K; $\ell^{\rm corr}$ used in all analyses.}
\label{tab:app_rept}\\
\toprule
Species & $M$ (kg) & $T_{\rm field}$ & $f_H^{\rm raw}$ & $f_H^{\rm corr}$ & $L$ (yr) & $\ell^{\rm raw}$ & $\ell^{\rm corr}$ \\
\midrule
\endfirsthead
\multicolumn{8}{c}{\tablename\ \thetable\ (continued)}\\
\toprule
Species & $M$ (kg) & $T_{\rm field}$ & $f_H^{\rm raw}$ & $f_H^{\rm corr}$ & $L$ (yr) & $\ell^{\rm raw}$ & $\ell^{\rm corr}$ \\
\midrule
\endhead
\bottomrule
\endfoot
\textit{Lacerta agilis}          & 0.015 & 301 &  45 &  93 & 12.0 & 8.45 & 8.77 \\
\textit{Anolis carolinensis}     & 0.006 & 302 &  52 & 106 &  6.0 & 8.22 & 8.52 \\
\textit{Pogona vitticeps}        & 0.350 & 303 &  42 &  82 & 10.0 & 8.34 & 8.63 \\
\textit{Phrynosoma cornutum}     & 0.035 & 301 &  48 &  99 &  7.0 & 8.25 & 8.56 \\
\textit{Iguana iguana}           & 4.000 & 303 &  40 &  79 & 20.0 & 8.62 & 8.92 \\
\textit{Varanus komodoensis}     & 65.0  & 303 &  28 &  55 & 30.0 & 8.65 & 8.94 \\
\textit{Tupinambis merianae}     & 2.500 & 302 &  38 &  77 & 15.0 & 8.48 & 8.78 \\
\textit{Thamnophis sirtalis}     & 0.050 & 300 &  30 &  62 & 10.0 & 8.20 & 8.51 \\
\textit{Coluber constrictor}     & 0.340 & 301 &  35 &  72 & 13.0 & 8.38 & 8.69 \\
\textit{Python reticulatus}      & 75.0  & 302 &  20 &  41 & 25.0 & 8.42 & 8.73 \\
\textit{Boa constrictor}         & 15.0  & 301 &  25 &  52 & 40.0 & 8.72 & 9.04 \\
\textit{Chelonia mydas}          & 180.0 & 300 &  20 &  42 & 80.0 & 8.93 & 9.25 \\
\textit{Geochelone gigantea}     & 200.0 & 298 &  15 &  33 &175.0 & 9.14 & 9.48 \\
\textit{Gopherus agassizii}      & 4.500 & 299 &  22 &  47 & 80.0 & 8.97 & 9.30 \\
\textit{Sphenodon punctatus}     & 0.800 & 293 &  18 &  43 & 77.0 & 8.86 & 9.24 \\
\textit{Crocodylus niloticus}    & 400.0 & 303 &  25 &  49 & 70.0 & 8.96 & 9.26 \\
\textit{Alligator mississippiensis} & 250.0 & 302 & 28 & 57 & 50.0 & 8.87 & 9.18 \\
\end{longtable}

\subsection*{A.8\quad Amphibians, Arrhenius-corrected ($n = 9$)}

Correction method identical to reptiles.
Clade corrected mean $\bar\ell^{\rm corr} = 8.822 \pm 0.146$.

\begin{longtable}{llrrrrrrl}
\caption{\textbf{Amphibians, Arrhenius-corrected.}}
\label{tab:app_amph}\\
\toprule
Species & $M$ (kg) & $T_{\rm field}$ & $f_H^{\rm raw}$ & $f_H^{\rm corr}$ & $L$ (yr) & $\ell^{\rm raw}$ & $\ell^{\rm corr}$ \\
\midrule
\endfirsthead
\multicolumn{8}{c}{\tablename\ \thetable\ (continued)}\\
\toprule
Species & $M$ (kg) & $T_{\rm field}$ & $f_H^{\rm raw}$ & $f_H^{\rm corr}$ & $L$ (yr) & $\ell^{\rm raw}$ & $\ell^{\rm corr}$ \\
\midrule
\endhead
\bottomrule
\endfoot
\textit{Rana temporaria}         & 0.025 & 294 & 25 & 55 & 16.0 & 8.32 & 8.67 \\
\textit{Rana catesbeiana}        & 0.500 & 296 & 20 & 43 & 16.0 & 8.23 & 8.56 \\
\textit{Bufo bufo}               & 0.150 & 293 & 22 & 53 & 36.0 & 8.62 & 9.00 \\
\textit{Xenopus laevis}          & 0.200 & 295 & 20 & 45 & 30.0 & 8.50 & 8.85 \\
\textit{Ambystoma mexicanum}     & 0.300 & 294 & 18 & 41 & 25.0 & 8.37 & 8.73 \\
\textit{Salamandra salamandra}   & 0.080 & 290 & 20 & 49 & 24.0 & 8.40 & 8.79 \\
\textit{Plethodon glutinosus}    & 0.012 & 291 & 30 & 74 & 20.0 & 8.50 & 8.89 \\
\textit{Necturus maculosus}      & 0.130 & 288 & 18 & 46 & 30.0 & 8.45 & 8.86 \\
\textit{Cryptobranchus alleganiensis} & 0.600 & 289 & 15 & 39 & 55.0 & 8.64 & 9.05 \\
\end{longtable}

\subsection*{A.9\quad Dataset summary}

\begin{table}[H]
\caption{\textbf{Summary statistics for all 230 species.}
$\bar\ell \pm s$: mean $\pm$ s.d.\ of $\ell = \log_{10}(f_H^{\rm eff} \cdot L \cdot 525{,}960)$.
$\Delta\bar\ell$: deviation from non-primate placental baseline ($\bar\ell_0 = 8.994$).
$\Phi = 10^{\Delta\bar\ell}$.
Significance vs baseline by Welch $t$-test:
$^*p < 0.05$, $^{***}p < 0.001$, ns = not significant.}
\label{tab:app_summary}
\small
\begin{tabular}{lrrrll}
\toprule
Group & $n$ & $\bar\ell \pm s$ & $\Delta\bar\ell$ & $\Phi$ & Sig. \\
\midrule
Non-primate placentals & 46 & $8.994 \pm 0.159$ & 0 (ref.) & 1.00 & --- \\
Marsupials/monotremes  & 19 & $8.933 \pm 0.209$ & $-0.062$ & 0.87 & ns \\
Primates               & 18 & $9.376 \pm 0.129$ & $+0.382$ & 2.41 & $^{***}$ \\
Bats (duty-corr.)      & 31 & $9.541 \pm 0.166$ & $+0.547$ & 3.52 & $^{***}$ \\
Cetaceans (dive-corr.) & 12 & $8.802 \pm 0.308$ & $-0.192$ & 0.64 & ns \\
Birds                  & 78 & $9.528 \pm 0.214$ & $+0.534$ & 3.42 & $^{***}$ \\
Reptiles (Arr.-corr.)  & 17 & $8.929 \pm 0.301$ & $-0.064$ & 0.86 & ns \\
Amphibians (Arr.-corr.)&  9 & $8.822 \pm 0.156$ & $-0.172$ & 0.67 & $^{*}$ \\
\midrule
\textbf{Full dataset}  & \textbf{230} & $9.253 \pm 0.355$ & & & \\
\bottomrule
\end{tabular}
\end{table}

\end{document}